\documentclass{IEEEtran}
\usepackage{cite}
\usepackage{amsmath,amssymb,amsfonts}
\usepackage{algorithmic}
\usepackage{graphicx}
\usepackage{textcomp}
\usepackage{color}

\def\BibTeX{{\rm B\kern-.05em{\sc i\kern-.025em b}\kern-.08em
    T\kern-.1667em\lower.7ex\hbox{E}\kern-.125emX}}
\begin{document}
\title{Electrically small matched antennas with time-periodic and space-uniform modulation}
\author{Alessio D'Alessandro \IEEEmembership{(*)}
\thanks{ (*) via Roggerone 3/6 16159 Genova, Italia (email:\mbox{ alessiorivarolo@yahoo.it).}}
}

\markboth{Submitted to IEEE Transaction on Antennas and Propagation, \textcopyright IEEE 2020}%
{*** \MakeLowercase{\textit{et al.}}: Bare Demo of IEEEtran.cls for IEEE Journals}

\maketitle

\begin{abstract}

A conceptual method to realize an impedance-matched electrically small antenna is discussed, using a transmission line modulated only in time. The time-periodic space-uniform modulation of a distributed shunt array of varcapacitors excites a mode with very small frequency but still a small (microwave) wavelength: the matching line then retains a practical size. The modulated transmission line brings the problem of impedance matching to a frequency $f_0$ much higher than the signal, where the large antenna reactance can be easily compensated with a much higher bandwidth for the same $Q$ factor: actually the bandwidth $\Delta f= f_0 / Q$ becomes large when $f_0$ is much larger than the signal frequency.
A simple way to obtain the space uniform modulation of the transmission line is illustrated, making use of a Wheatstone-bridge arrangement of the modulated varicaps. The signal and modulating lines, running in parallel on a section of the same length, see two very different equivalent shunt capacitances, so that the modulating voltage is almost uniform while the signal voltage may change significantly across the transmission line.

\end{abstract}


\section{Introduction}
\label{sec:introduction}
Getting radiation from an electrically small antenna, an antenna much shorter than the radiation wavelength,
has been a longstanding problem in radio engineering since early times.
At low frequencies the radiation resistance of a small dipole scales as the square of the ratio of the antenna and radiation characteristic lengths, which can be extremely small. Moreover, besides the dramatic reduction of the radiation resistance, the impedance of an electrically small dipole antenna has a huge capacitive component and poses well-known matching problems with the source.

Compensation of the huge capacitive component with passive linear time-invariant (LTI) elements (i.e. a large inductance) can occur just in a very narrow frequency bandwidth dictated by the Bode-Fano limit \cite{limit1,limit2} and it is impractical for electrically small antennas; furthermore, the resistive losses of the (non-ideal) inductive compensator can be much larger than the antenna radiation resistance. 
That is why alternative impedance matching methods have been deeply investigated in literature, like non-Foster active networks \cite{nonfoster1} with a negative effective capacitance, negative refraction index metamaterials \cite{doubleneg1,doubleneg2,doubleneg3}, original meander-like antenna shapes \cite{meander} and also magnetoelectric antennas \cite{acoustic1,acoustic2,acoustic3,acoustic4,acoustic5, hadad}.

Nonlinear, modulated time variant networks (see for instance \cite{switchshort,switchline,modulatedC}) are a further possible way to match the impedance of electrically small antennas without the bandwidth limitation of Bode-Fano for LTI systems. Moreover, in recent times, space-time modulation of metasurface structures has been attracting a rise of interest in many other contexts beside small antennas, such as parametric amplifiers, multichannel duplexers, and optical insulators (i.e. reflectionless EM absorbers) \cite{metasurface1,metasurface2,metasurface3,leakantenna,opticaltrans,graphene}. 

In this paper, the space-uniform modulation method is applied to get a time-variant $\lambda/4$ stub section of transmission line, as an impedance adapter for an electrically small antenna. As for any adapter, the impedance seen at the generator terminals before the $\lambda/4$ section is different from the impedance seen at the antenna terminals and can be matched much more easily. 

A $\lambda/4$ section of transmission line is not trivially realizable at low frequencies, and normally needs complex meander structures \cite{meander} to realize enough length in a (still) practical space: here modulation comes into help and dramatically reduces the needed size.

The modulation of the fundamental high frequency signal $f_G$ of the generator (input) with a space-uniform signal of a very close frequency (modulation) provides a low frequency (\emph{beat}) excitation \emph{with the same small wavelength of the generator} (since the modulation is uniform in space) and frequency equal to the difference of the fundamental and modulation: this is the low frequency $f_s$ emitted by the small antenna. 

Thanks to the space-uniform modulation, the $\lambda/4$ matching element propagates a signal with the very low antenna frequency and the practical-size small wavelength of the generator side at the same time. As a consequence, the significant length $\lambda/4=c/(4 f_G)$ needed for the impedance matching is defined \emph{at the higher generator frequency $f_G$}, and is much smaller than $c/(4 f_s)$.

The idea of applying a modulated lumped load to an electrically small antenna has already been theorized recently \cite{modulatedC}: in receiving mode it is claimed to enhance the bandwidth by a factor of two compared to the situation without modulated load. The more complex method of space-uniform modulation is here not applied to a lumped element but to a whole distributed system (transmission line) in transmission mode to obtain a larger bandwidth, as we will see.

We point out that the present method does \emph{not} affect the antenna characteristic impedance seen by its terminals, which is unchanged and always poor. Instead we just provide an upstream (modulated) line that facilitates the compensation of the antenna reactance with appropriate elements at high frequency. 

This operation can be done with a significant larger bandwidth compared to the standard matching technique at a single frequency. 
We will show that for the same value of $Q$ factor, the bandwith defined at the high frequency $f_0$ is $\Delta f=f_0/Q$, that is larger of a factor $f_0/f_s$.
This is the ultimate feature that makes the matching at the modulated (high) frequency much more appealing than the matching at the signal frequency.

This paper is structured as follows: in Section \ref{sec:modulation} the main idea of time periodic and space-uniform modulation for electrically small antennas is sketched; in Section \ref{sec:antenna} a system which can provide the needed modulation is discussed; in Section \ref{sec:boundaries} the boundary conditions (filters), in Section \ref{sec:bandwidth} the bandwidth and, finally, in Section \ref{sec:limitations} the limitations of the method are illustrated.

\section{Time periodic modulation for small antennas}
\label{sec:modulation}

A wave of angular frequency $\omega_0$ propagating in a conductor with a propagation speed modulated in time, but uniform in space,
$c(t)=c/(\sqrt{\epsilon}(1+\delta \cos(\omega_m t)))$, where $\epsilon$ is the dielectric permittivity and $\delta$ the modulation index, can excite modes of different angular frequencies $\omega_n = \omega_0 + n \omega_m$ with the same spatial wavenumber $\sqrt{\epsilon}\omega_0/c$.
This appears intuitively from the na\"{i}ve first order Taylor expansion 
\begin{IEEEeqnarray}{l}
\exp\left(i \left(\frac{\omega_0  }{c(t)} z - \omega_0 t \right)\right)  =  
\exp\left(i \left(\frac{\sqrt{\epsilon} \omega_0  }{c} z - \omega_0 t \right)\right) - \nonumber \\
- \sum_{n=1,-1}\frac{i \delta \sqrt{\epsilon} \omega_0 z}{2 c}\exp\left(i \left(\frac{\sqrt{\epsilon} \omega_0  }{c} z - (\omega_0 + n \omega_m) t \right)\right)
\label{eq}
\end{IEEEeqnarray}
for a small modulation index $\delta$.

\begin{figure}[!t]
\centerline{\includegraphics[width=\columnwidth]{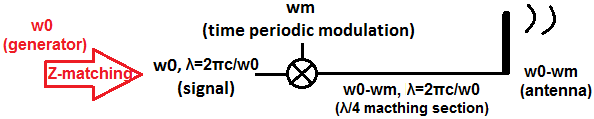}}
\caption{Schematic of the method: the time-periodic modulation of a high frequency signal at $\omega_0$ induces a low frequency signal at $\omega_0-\omega_m$ transmitted by the antenna. The generator at high frequency sees an adapted impedance (different than the one at the antenna terminals) thanks to the $\lambda/4$ matching line.
}
\label{diagram}
\end{figure}

The idea behind this paper is to provide a (space-uniform) modulation across a $\lambda/4$ section of matching transmission line, with a modulation (angular) frequency $\omega_m$ very close to the base radiofrequency 
$\omega_0$. A low frequency wave with small frequency $\omega_0-\omega_m<<\omega_0$
is excited, with the same fundamental spatial wavenumber as the generator side (Figure~\ref{diagram}). 

\section{The antenna system}
\label{sec:antenna}
\begin{figure}[!t]
\centerline{\includegraphics[width=\columnwidth]{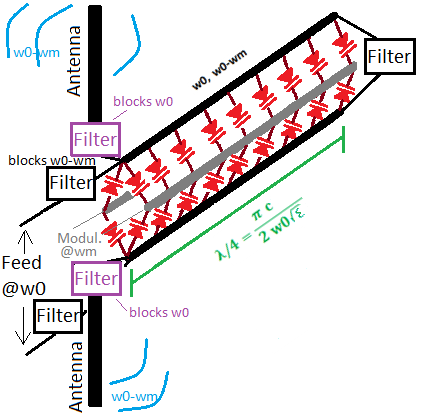}}
\caption{Overview of the $\lambda/4$ matching stub to the electrically small antenna, where
$\lambda/4=\frac{\pi c}{2 \sqrt{\epsilon}\omega_0}$ is defined with respect to the feed frequency $\omega_0$ and not with the much smaller signal frequency $\omega_0-\omega_m$.
The pair of conductors in bold black is reserved for the feed/signal frequency while the other couple of parallel conductors, in grey, carry the modulation signal, (almost) uniform in space across the transmision line length. Alternative options for the cross section are provided in Figure \ref{frontview}.
The antenna and feed are placed on the same side of the transmission line with a system of filters depicted in Figure \ref{sideview}.  
 Inessential dielectric embedding (assumed $\epsilon_r=1$) is not shown. Assuming a minimum 5~mm spacing between the varicap rows as in 
\cite{leakantenna} and a minimum number of $N=20$ rows for the continuum limit (\ref{telegraph}) to be realized, the $\lambda/4$ matching line should be longer than 10~cm.}
\label{antenna}
\end{figure}

Figure \ref{antenna} shows an example of space-uniformly modulated $\lambda/4$ matching stub, where $\lambda$, as said, is defined with the (high) generator input frequency $\omega_0$ and not with the (small) antenna signal frequency $\omega_0-\omega_m$.

Time-periodic space-uniform modulation is not as simple as space-time modulation. While for the latter a simple travelling wave on a dedicated line can be used to modulate the gap voltage of an array of varactors, like in \cite{leakantenna}, space uniform modulation requires a method for synchronizing the phase of the modulation wave across the length of the system: for a given time instant, each varactor must receive its modulating voltage with the same phase. 

\begin{figure}[!t]
\centerline{\includegraphics[width=\columnwidth]{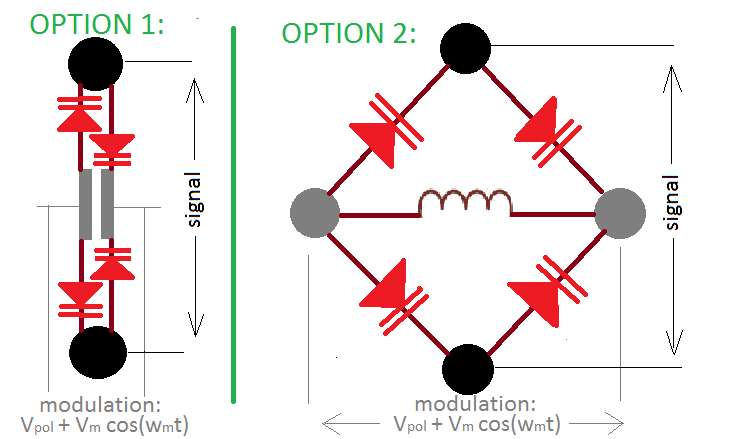}}
\caption{Front section of Figure \ref{antenna} displaying two alternative Wheatstone bridge arrangements of the varicaps. For OPTION 1, the modulating line (in gray) has an unloaded shunt capacitance $C_{line}$ much greater than the one of the signal line (in black), since gray conductors are closer and wider: for the modulation line the extra capacitance of the varicap array is negligible.  For OPTION 2, the compensating inductance is seen only by the modulating line due to the Wheatstone bridge arrangement. 
}
\label{frontview}
\end{figure}

A practical way to obtain (an approximate) space uniform modulation across the $\lambda/4$ stub of Figure \ref{antenna} is to arrange the varicap array in a Wheatstone bridge configuration as in Figure \ref{frontview}, with two alternative options as cross section. 

As first option of Figure \ref{frontview} (left), a very different spacing and width is assigned to the conductors of the modulation line and the signal line, the former being much closer and wider and, consequently, having a much greater shunt capacitance than the latter: the extra capacitive load of the varicap array is negligible for the modulation line and the propagation speed of the modulating signal is $c$.
For the signal line, instead, the conductor separation and size is designed so that the varicap capacitance is greater than the unloaded line capacitance of a factor $\epsilon>>1$: this sets a much smaller phase velocity $c/\sqrt{\epsilon}$, and then a much smaller wavelength than the modulation line. With this arrangement the $\lambda/4=\frac{\pi c}{2 \sqrt{\epsilon}\omega_0}$ section of transmission line in Figure \ref{antenna} can be much smaller of the characteristic quarter-wavelength $\frac{\pi c}{2 \omega_m}\sim\frac{\pi c}{2 \omega_0}$ of the modulating voltage, which is then approximately constant across the whole length. 

The second option in order to assign a much larger characteristic wavelength to the modulating voltage than to the signal voltage is to use a compensating inductance array (Figure \ref{frontview} right), resonant with (the unmodulated part of) the varicap capacitance at the modulation frequency.
The signal line and the modulating line are then realized with parallel conductors lying on orthogonal planes: the equivalent capacity across the signal line is 
the sum of the (modulated) shunt capacitance of the varicap array and the line capacitance of the parallel conductors, since, when the Wheatstone bridge is balanced, there is no signal current across the inductance.
On the other hand, the equivalent capacitance seen across the modulating line is just the conductor line capacitance, since (the unmodulated part of) the varicap capacitance is compensated by the shunt inductance. If we choose the varicap capacitance to be much larger than the transmission line capacitance, then the wavelength associated to the signal line $2\pi/(\omega_0\sqrt{L (C_{line}+C_{varicap})})$ is much smaller than the wavelength of the modulating line $2\pi/(\omega_m\sqrt{L C_{line}})$, where $\omega_m\sim\omega_0$ is the modulation frequency and $\omega_0$ the feed frequency. Although the conductors of the signal and the modulating lines run in parallel with the same length, only for the latter this lenght is electrically small and yields (almost) uniform modulating voltage across the whole length. 

For both options of Figure \ref{frontview}, the high symmetry of the Wheatstone bridge configuration makes the varicap capacity highly independent of the feed voltage of the signal: in a back-to-back configuration, any change of the signal voltage induces equal and opposite changes in the capacity of each back-to-back condenser and the overall series reactance $\frac{1}{j \omega (C +\delta C)}+\frac{1}{j \omega(C -\delta C)}=\frac{1}{j \omega C}+o(\delta C/C)$ stays unchanged up to second orders in the modulation index. The high symmetry also prevents any undesired coupling of the signal/feed line with the modulation line: noise in the modulation line affects in the same way each conductor of the signal line, thus producing no differential voltage across the signal.

The two conductor bars of the signal line are connected to a high frequency voltage feed, providing the fundamental wave at frequency $\omega_0$, and to the (electrically small) antenna on the same end, through a system of filters shown in Figure \ref{antenna} and \ref{sideview} and discussed later on. The other couple of conductors, the modulation line, is connected with a modulating source at a frequency $\omega_m$ close to $\omega_0$, and provides a nearly uniform voltage in space across the length of the rods. The DC polarization circuit can be provided on the same side of the modulation feed with a decoupling condenser as in Figure \ref{sideview}.

For our simplified analysis we assume the matching system to be embedded in a lossless dielectric with $\epsilon_r=1$ with mere function
of support. We also neglect the varicap parasite resistance: even though the single varicap element may have a parasite resistance ($\sim1.8\Omega$ for Infineon BB833) of the same order as the radiation resistance of the small antenna, the equivalent resistance of the parallel arrangement is much smaller: this may be an appreciable feature of the space-uniform distributed modulation compared to the lumped modulation.

A capacitance (per unit length) of the varicap array $C_{0\ varicap}$ much higher than the  natural capacitance $C_{line}$ of the signal parallel line provides a high effective dielectric permittivity $\epsilon=C_{0\ varicap}/C_{line}$: this in turn allows a shorter length 
$\frac{\lambda}{4}=\frac{\pi c}{2 \sqrt{\epsilon}\omega_0}$ of the conductor rods for a given frequency $\omega_0$, with consequent space saving. 

With Infineon BB833 varicaps polarized at 2~V with a corresponding 7~pF capacitance, a 5~mm spacing of the varicap rows as in \cite{leakantenna} allows to have $C_{0\ varicap}$=7~pF/(5mm), a much larger value than the unloaded line capacitance of $C_{line}$=9.3 pF/m for round conductors spaced 20 times the radius. For this configuration $\sqrt{\epsilon}\sim$12.3 and with $\lambda/4>10$cm, corresponding to a number $N>20$ of varicap rows, the condition of space-uniform modulation is well achieved. 

On the same side as the antenna, the couple of conductor bars carrying the signal is fed with differential voltage $V(0,t)=V_0 \exp(-i \omega_0 t)$, with the fundamental $\omega_0$ in the UHF frequency range. A fraction of the generated power at $\omega_0$ is converted into power at $\omega_0-\omega_m$ by the modulated varicaps as long as the signal travels along the line, while the remaining fraction enters the modulation circuit. At the opposite end of the antenna, pure capacitive filters set a zero power boundary condition at both frequencies $\omega_0$ and $\omega_0-\omega_m$, as we will see in Section \ref{sec:boundaries}, ensuring that all the feed energy at $\omega_0$ has been converted. It will be shown that energy at $\omega_0-\omega_m$ travels the opposite direction with respect to energy at $\omega_0$ along the modulated line, that is toward the antenna.

The telegraphers' equations 
\begin{IEEEeqnarray}{rcl}
\partial_z I(z,t) &=& - \partial_t (C(t) V(z,t)) \nonumber \\
\partial_z V(z,t) &=& - L\ \partial_t I(z,t) 
\label{telegraph}
\end{IEEEeqnarray}

decribe the propagation of a voltage $V(z,t)$ and current $I(z,t)$ wave between the two parallel conductor bars loaded with the time-modulated capacitor array, in parallel with their natural (unmodulated) capacitance. The capacitance per unit length $C(t)=C_{line}+C_{0\ varicap}(1+\delta \cos(\omega_m t))$ includes both the unmodulated transmission line capacity $C_{line}$ (for round conductors of radius $a$, at a distance $d$ apart, $C_{line}=\frac{\pi \epsilon_0}{\log(d/a)}$) and the varicap capacitance as well, the latter sized to be much larger than the line capacitance. 
In the above equations, $d$ is the separation gap between conductors and $z$ is the coordinate along the transmission line.

Neglecting dielectric leakage losses and conductor resistance, equations (\ref{telegraph}) can be rearranged into the wave equation:
\begin{equation}
\partial_z^2 V(z,t) - L\ \partial_t^2 ( C(t) V(z,t)) =0  
\label{waveeq}  
\end{equation}
It can be shown (Appendix \ref{sec:wavesolution}) that the equation above has solution for modes with frequency $\omega_0 + n \omega_m$ for $n$ integer: the low frequency mode $\omega_0 - \omega_m$ is combination of a wave with short wavelength $\lambda=\frac{2 \pi c}{\sqrt{\epsilon} \omega_0}$ generated by the modulation, which allows the impedance matching, along with the standard wave with long wavelength $\lambda=\frac{2 \pi c}{\sqrt{\epsilon} (\omega_0-\omega_m)}$ that would be there even in absence of modulation.   
 
Boundary conditions are carefully chosen to suppress (angular) frequencies different from $\omega_0$ and $\omega_0-\omega_m$, thanks to additional dedicated filters at the conductor ends, as discussed in Section \ref{sec:boundaries}. 
Propagation of modes $\omega_n=\omega_0 + n \omega_m$ with $n\neq0,-1$ is also discouraged by the smallness of the modulation index $\delta$ 
($n$-mode amplitudes are proportional to $\delta^{|n|}$).

\section{Boundary conditions and filters \label{sec:boundaries}}

A full set of voltage and current solutions of the modulated wave equation (\ref{waveeq}) is derived in Appendix \ref{sec:wavesolution} (expressions 
(\ref{Vmodesol}) and (\ref{Imodesol})): they are linear combination of a "short" wave with the natural wave number $\beta_0=\frac{\sqrt{\epsilon} \omega_0}{c}$ of the feed frequency $\omega_0$ (i.e. the wave number that the feed frequency would have in absence of modulation), and a "long" wave with the natural wave number $\beta_{-1}=\frac{\sqrt{\epsilon} (\omega_0-\omega_m)}{c}$ of the radiated signal frequency. Both modes ($\omega_0$ and $\omega_0-\omega_m$) carry both wavelengths, in shares defined by the boundary conditions: the latter will fix the generic (voltage-dimensional) coefficients $A,B,C,D$ of the linear combination in (\ref{Vmodesol}) and (\ref{Imodesol}). The "interesting" part is of course the "short" wave contribution associated to the low frequency mode, that is proportional to the modulation index $\delta$ and would be zero in absence of modulation: the antenna impedance matching is made possible thanks to this part.

Figure \ref{sideview} shows the boundary conditions set by the filters: filters F0$_{out}$ and F1$_{out}$ must be pure reactive elements. Their impedance at the corresponding operating frequency is
F0$_{out}=Z_0 \frac{2\omega_0^2}{i\pi(\omega_0-\omega_m)^2}$ for the filter at $\omega_0$ and F1$_{out}=Z_0 \frac{2\omega_0}{i\pi(\omega_0-\omega_m)}$ for the filter at $\omega_0-\omega_m$, as calculated in the Appendix \ref{sec:wavesolution}, where $Z_0=\sqrt{L_{line}/C_{0\ varicap}}=\sqrt{L_{line}/(\epsilon\ C_{line})}$. They can be realized with simple condensers in series with frequency selective filters. 

At the antenna end, the block-band filter F0$_{in}$ (realized with a standard LC parallel arrangement tuned to the blocking frequency $\omega_0$) ensures that the feed frequency $\omega_0$ does not enter the antenna, which should radiate the low mode $\omega_0-\omega_m$ only. At the antenna terminals it is really advisable to use a block-band filter F0$_{in}$ (LC parallel) rejecting the unwanted frequency $\omega_0$ rather than a pass-band filter (LC series) selecting the low mode $\omega_0-\omega_m$. Actually, we do not want to reduce the bandwidth by creating a not needed resonance at $\omega_0-\omega_m$ with the filter: \emph{the large antenna reactance is to be compensated at $\omega_0$, where the bandwidth is large, not at $\omega_0-\omega_m$}. Moreover, a block-band LC parallel arrangement at high frequency, unlike the LC series bandpass at low frequency, does not introduce a high parasite resistance of the compensating inductor which could be of the same order as the low radiation resistance of the antenna.

The remaining filter F1$_{in}$ is a LC series bandpass tuned to $\omega_0$,  which ensures that the power at signal frequency $\omega_0-\omega_m$ is not dissipated in the source resistance of the feed. Finally, filters F2 are band-pass filters shorting out the high frequency mode $\omega_0+\omega_m\sim2\omega_0$: since the wavelength associated to this mode is roughly one half the wavelength at $\omega_0$, an additional F2 filter is placed mid-way the transmission line to remove the occurrence of the stationary wave 
at $\omega_0+\omega_m$ with nodes at the end of transmission line and peak in the middle, which could not be removed by the filters F2 at the bound of transmission line.

\section{Bandwidth and power}
\label{sec:bandwidth}
\begin{figure}[!t]
\centerline{\includegraphics[width=\columnwidth]{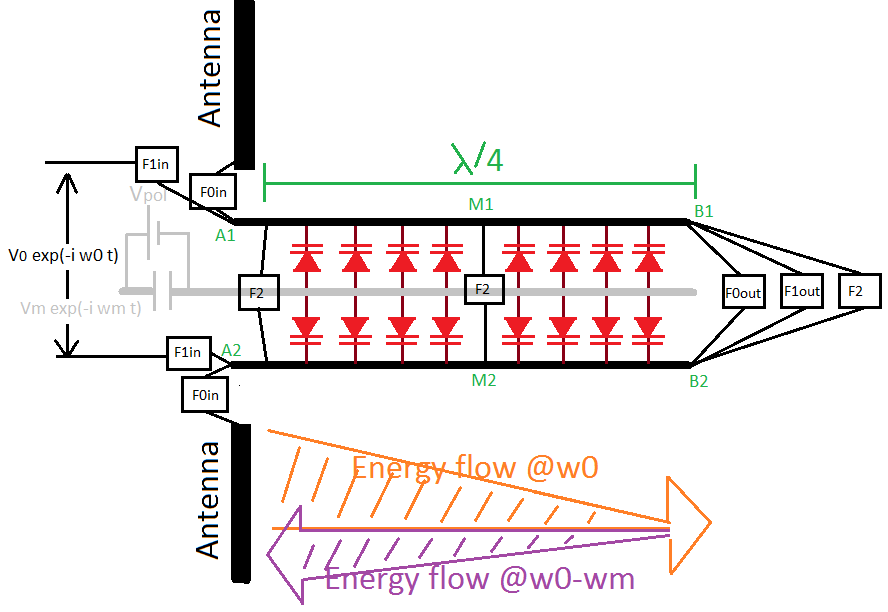}}
\caption{Side view of the $\lambda/4$ matching system to the electrically small antenna with modulation line, filters and polarization circuit. The power flow at $\omega_0$ decreases from a positive value at the feed terminals A1 and A2 to zero at the end opposite the antenna as long as energy is introduced in the modulation circuit. Power at the low frequency mode, introduced by the modulated varicaps, travels the opposite direction (toward the antenna) from the zero value at the terminals B1 and B2 to the maximum value at the antenna 
terminals. 
}
\label{sideview}
\end{figure}

With the assessment above, the input impedance at $\omega_0$, at the feed points A1 and A2 in \mbox{Figure \ref{sideview}}, is 
$Z_{w0\ in}\sim Z_{ant}\frac{(\omega_0-\omega_m) \delta^2}{4\omega_0}$ as calculated in (\ref{Z0inputsimpl}) in the Appendix.
This impedance can be easily compensated from the feed circuit upstream points A1 and A2, with a small inductor of opposite reactance
$\Im(Z_G)=-\Im(Z_{ant})\frac{(\omega_0-\omega_m) \delta^2}{4 \omega_0}$ at $\omega_0$, cancelling the imaginary part. For the real part, either a small-size high-frequency transformer or a short matching section of unmodulated transmission line at $\omega_0$ can be used to set $\Re(Z_{ant})\frac{(\omega_0-\omega_m) \delta^2}{4 \omega_0}$ equal to the generator resistance $\Re(Z_G)$. 

At this point it is useful to compare the radiated power at the low frequency $\omega_0-\omega_m$ with the feed power entering the 
modulation system (at terminals A1 and A2 in Figure \ref{sideview}) at $\omega_0$.
The provided feed power is \mbox{$P_{0}=\Re(Z_{w0\ in}) |I_{0}(0)|^2$} while the power at the antenna terminals is $P_{-1}=\Re(Z_{w1\ in}) |I_{-1}(0)|^2=-\Re(Z_{ant}) |I_{-1}(0)|^2$
(negative in sign since power flows from the line into the antenna), where the moduli $|I_{0}(0)|$ and $|I_{-1}(0)|$ of the $z=0$ input currents at each frequency are provided by equation (\ref{Imodesol}) evaluated at $z=0$ with the condition $D=0$ enforced by the boundaries (\ref{ABCD}).
We obtain $|I_{-1}(0)|/|I_{0}(0)|=\frac{\omega_0 (\omega_0-\omega_m) \delta }{2 \omega_m  (2 \omega_0-\omega_m) }\sim \frac{(\omega_0-\omega_m) \delta }{2 \omega_0 }$ so that $P_{-1}/P_{0}=-(\omega_0-\omega_m)/\omega_0$. This is a well expected result by the Manley-Rowe theorem
\cite{MRowe1,MRowe2}, which states that that for each power contribution flowing into the modulated varactors at $\omega_0$ there is one at $\omega_0-\omega_m$ opposite in sign and smaller of a fraction $\frac{(\omega_0-\omega_m)}{\omega_0}$. Since varactors are pure reactive elements, the balance fraction of energy
$1-\frac{(\omega_0-\omega_m)}{\omega_0}$ is not lost, but enters the modulation circuit at frequency $\omega_m$ where it could be either recovered for other purposes or eventually dissipated in the source resistance of the modulation signal.
Manley-Rowe theorem sets the conversion rate $\frac{(\omega_0-\omega_m)}{\omega_0}$ of the feed power, no matter how the antenna is realized and regardless of the boundary condition: by taking the time-averaged power $\frac{1}{2}\Re(V(z,t)I^*(z,t))$ from (\ref{Vmodesol}) and (\ref{Imodesol}) it can be also proven that $\partial_z P_{0}(z) = - \frac{\omega_0}{(\omega_0-\omega_m)} \partial_z P_{-1}(z)$ holds at a generic $z$ for any value of the boundary condition coefficients $A,B,C,D$.

At the feed terminals A1 and A2, power at $\omega_0$ is positive, that is entering the modulated line, while power at $\omega_0-\omega_m$ is negative, 
that is coming out from the modulated line (into the antenna); at the end $z=\lambda/4$ opposite the antenna there is no power flow at neither of the two modes, due to the pure reactive boundary conditions: all the feed power $P_{0}$ has been converted by the varactors. The minus sign in the
Manley-Rowe relationship above enforces the power flow at frequency $\omega_0-\omega_m$ in the opposite direction of $\omega_0$, as shown in Figure \ref{sideview}.
The power transmitted to the antenna is then a fraction 
$(\omega_0-\omega_m)/\omega_0$ of the feed power sent to the transmission line, which, due to the impedance matching at $\omega_0$ before the terminals A1 and A2, is one half of the generated power at $\omega_0$ (the other half lost in the generator resistance).

From a first impression it may look deceptive that only a fraction $\frac{(\omega_0-\omega_m)}{\omega_0}$ of the feed power is converted into the low frequency to be radiated by the antenna, but a closer look shows that it is not so bad. Actually we obtained a performance \emph{linear} in the small frequency, while for the generic case of an unmatched antenna the radiated power is only a fraction $R_{ant}/R_G$, where $R_G$ is the source resistance and $R_{ant}$ is the radiation resistance: $R_{ant}$ and the ratio $R_{ant}/R_G$ are \emph{quadratic} in the small frequency, hence much smaller.
Moreover, while in the latter case the residual power is entirely dissipated in the source resistance, in the present case of a modulated matching line the balance of power is just \emph{converted} to a different frequency $\omega_m$. In other words, the modulated line works as a frequency parametric downconverter where the idler circuit operates at a frequency $\omega_0-\omega_m$, much lower than the feed $\omega_0$ and than the pump (modulation) frequency $\omega_m$.
This result is not only a performance improvement (linear behavior in the signal frequency, instead of quadratic behavior) but also a bandwidth improvement with respect to any "standard" low frequency matching.

The Q factor of the match at $\omega_0$ can be calculated from \cite{Qfactor}: 
\begin{IEEEeqnarray}{rcl}
Q&=&\frac{\omega_0 |\Im(Z_{w0\ in}'(\omega_0)+Z_G'(\omega_0))|}{2 \Re(Z_{w0\ in}(\omega_0))}=\frac{|\Im(Z_{w0\ in}(\omega_0))|}{\Re(Z_{w0\ in}(\omega_0))}=\nonumber \\
&\sim&\frac{|\Im(Z_{ant})|}{\Re(Z_{ant})}
\end{IEEEeqnarray}
where $\Im(Z_G)$ is the compensating reactance of the feed. In the derivation of the RHS of the above we assume the canonical
inverse-frequency dependence for the reatance of the electrically small dipole antenna, that is $\Im(Z_{ant})=-\frac{1}{(\omega_0-\omega_m) C_{ant}}$ so that 
$\Im(Z_{w0\ in}(\omega_0))=-\frac{\delta^2}{4 \omega_0 C_{ant}}$ and the compensating reactance is of the form $\Im(Z_G(\omega))=\frac{\omega \delta^2}{4 \omega_0^2 C_{ant}}$ (which can be obtained with a small inductor of
value $\frac{\delta^2}{4 \omega_0^2 C_{ant}}$). In this way $\Im(Z_{w0\ in}'(\omega_0))= -\Im(Z_{w0\ in}(\omega_0))/\omega_0$ and $\Im(Z_{G}'(\omega_0))= \Im(Z_{G}(\omega_0))/\omega_0 = -\Im(Z_{w0\ in}(\omega_0))/\omega_0$.
The Q factor is then the same as a conventional matching at $\omega_0-\omega_m$, but the bandwidth at $\omega_0$, defined as $\Delta \omega= \omega_0/Q$, is enhanced by the (large) factor $\omega_0/(\omega_0-\omega_m)$ with respect to the conventional match at the signal frequency $\Delta \omega= (\omega_0-\omega_m)/Q$.

As a comparison, the most performant case of standard linear time invariant antenna model in \cite{Qcomp1}, that is the folded spherical helix, for the investigated case $k a=0.263$ ($a$=radius of Chu's sphere enclosing the antenna), has a Q factor of 84.78, which is close to the Chu limit $Q_{min}=\frac{1}{(k a)^3}+\frac{1}{k a}=58.8$: a slightly higher value around $Q=120$ is obtained in \cite{Qcomp2} for a spherical helix with less windings. 
A simple dipole antenna with the same $k a =0.263$ has length $L= 2 a$ and hence \mbox{$\Re(Z_{ant})\sim1.4\Omega$} and
\mbox{$|\Im(Z_{ant})|\sim 2$k$\Omega$} slightly depending on the radius of conductor. With the technique of space-uniform modulation, the large 
$Q_{dipole}=|\Im(Z_{ant})|/\Re(Z_{ant})\sim1400$ of the simple dipole could generate as much bandwidth as the complex spherically-folded geometry close to the Chu limit with a modulation frequency equal to $Q_{dipole}/Q_{min}=24$ times the signal frequency. The space-uniform modulation can be applied to a more complex antenna geometry than a dipole or to other standard methods enhancing the radiation resistance (e.g. capacitor end plates) by combining the techniques. 

\section{Limitations of the technique}
\label{sec:limitations}
An upper limit to the high frequencies $\omega_0$ and $\omega_m$ is set by the space-uniform nature of the modulation, which requires a minimum number of components (e.g. N=20 rows) to realize the continuum limit. With a minimum row spacing of 5mm set by the manufacturability (by simple welding), a 10cm minimum length for the $\lambda/4$  modulation line seems reasonable, which sets the maximum frequency $\omega_0<\frac{\pi c}{2\sqrt{\epsilon}\mbox{10cm}}$. 

A further, far more important limitation is on the ratio $\omega_0/(\omega_{0}-\omega_m)$ due to (\ref{Z0inputsimpl}):  not only the antenna reactance but also the small antenna resistance is seen smaller of a factor $\frac{(\omega_{0}-\omega_m)\delta^2}{4 \omega_0}$ on the high frequency side; for high values of $\omega_0/(\omega_{0}-\omega_m)$ it may fall below the parasite resistance of the matching element (e.g. transformer) with the generator source resistance. This limits the bandwidth enhancement, also proportional to $\omega_0/(\omega_{0}-\omega_m)$. For this reason, the space-uniform modulation technique, with the present settings (stub of length $\lambda/4$, output filters as in (\ref{Zoutputsimpl})), is more suitable to small antennas with an enhanced radiation resistance, such as multi-winding loop antennas. 

At present, although the theoretical basis here presented are encouraging, a simulation of a complete setup of antenna + filters + matching line has not been carried out yet and should be performed as a final validation of the practicity and applicability of the method.

\section{Conclusion}
The idea of time-periodic space-uniform modulation to provide a suitable low frequency excitation to electrically small antennas is sketched in a concept form. A high frequency voltage feed applied to a $\lambda/4$ stub section of transmission line with modulated 
line impedance generates a low frequency mode, with the same fundamental (microwave) wavelength, which can be used to transfer the impedance matching problem of the antenna to a much higher frequency $f_0$ than the signal $f_s$. At this frequency, the antenna impedance matching can be performed easily (e.g. by a small compensating inductor with negligible parasite resistance and a section of unmodulated transmission line of small length) and the bandwidth is enhanced by a factor $f_0/f_s$
(with the same $Q$-factor).

The method is here presented from a theoretical point of view, still needing a simulation of the complete arrangement to prove its practicity.

The same technique of space-uniform modulation with a high frequency $\omega_m$ close to the fundamental frequency $\omega_0$ of the feed can be possibly applied directly on the electrically small antenna instead of its matching circuit. A slot-loop antenna arrangement such as \cite{shuntC} can be loaded with an array of varicaps in the place of the simple capacitors used in \cite{shuntC}. 

It is worth saying that the large power fraction $1-\frac{\omega_0-\omega_m}{\omega_0}$ entering the modulation circuit 
due to the frequency downconversion is not necessarily dissipated in the source resistance of this circuit, but, in principle, can be recovered and eventually reconverted to $\omega_0$ by an additional parametric amplifier before the impedance matching circuit at $\omega_0$. 
The ultimate configuration should be a power feed at the low mode $\omega_0-\omega_m$ first converted to the high frequency mode $\omega_0$ with a parametric amplifier. Then an unmodulated line section at $\omega_0$ where the impedance match is performed before terminals A1 and A2 of Figure \ref{sideview}. Finally the modulated line downconverts the signal back to the low frequency $\omega_0-\omega_m$ to be radiated by the antenna.
The modulation for the upconversion and downconversion is provided by the same voltage so that the signal power employed in the downconversion is not lost, but recovered in its upconversion: two parametric converters (one lumped, one distributed, the modulated line) are in series to achieve unit power gain
$\frac{\omega_0}{\omega_0-\omega_m}\cdot\frac{\omega_0-\omega_m}{\omega_0}=1$.

This paper may also provide, in principle, a mathematical framework for the (technically challenging) generation of modulated signals in the teraHertz band \cite{opticaltrans}. A GHz feed signal of frequency $f_0$ could be theoretically upconverted by a space-uniform optical modulating signal (pump) to a much higher optical frequency $f_s$. In this case a huge power gain $f_s/f_0$ with $f_s>>f_0$ is predicted by the Manley-Rowe theorem, since the modulating part would act as parametric upconverter.

The above further investigations are eventually open to collaboration, in the awareness that no definite solution for 
a longstanding problem is provided here but just a first sketch of a method which may be gathered by others.

\appendices

\section{Solution of the modulated wave equation}
\label{sec:wavesolution}
A general voltage solution to the telegrapher's equation (\ref{waveeq}), for a couple of parallel conductors loaded with a time-periodic space-uniform modulated capacitance $C(t)$, evenly distributed along the conductors' length, is here presented.

The solution can be decomposed in the superposition of a forward and backward propagating wave, which in turn can be written in the Floquet expansion form:
\begin{IEEEeqnarray}{rcl}
V^\pm(z,t)=\exp(\pm i \beta z) \sum_n V_n \exp(-i \omega_n t) 
\label{ansatz}
\end{IEEEeqnarray}
with $\omega_n = \omega_0 + n \omega_m$. Here the $+$ sign is for the forward and the $-$ sign for the backward wave, and the pedix $n$ is the mode label. The mode amplitudes $V_n$ may have different values for the forward and the backward wave.

With $\epsilon:=C_{0\ varicap}/C_{line}>>1$ equation (\ref{waveeq}) becomes

\begin{IEEEeqnarray}{rcl}
\partial_z^2V(z,t) =  \frac{\epsilon}{c^2}\ \partial_t^2( (1+ \delta \cos(\omega_m t)) V(z,t))
\label{modesV0}
\end{IEEEeqnarray}

Plugging the ansatz (\ref{ansatz}) in the above equation we follow similar steps as \cite{leakantenna} to get:
\begin{IEEEeqnarray}{l}
\sum_n V_n \left[\left(-\beta^2 + \frac{\epsilon \omega_{n}^2}{c^2} \right) e^{-i \omega_{n} t}+\frac{\delta \epsilon \omega_{n+1}^2}{2 c^2} e^{-i \omega_{n+1} t}+ \right. \\
\left. +\frac{\delta \epsilon \omega_{n-1}^2}{2 c^2} e^{-i \omega_{n-1} t} \right]=0 \nonumber
\end{IEEEeqnarray}

Relabelling the index of terms under summation ($n\rightarrow n\pm1$) we obtain
\begin{IEEEeqnarray}{l}
\sum_n  e^{-i \omega_{n} t} \left[\left(-\beta^2 + \frac{\epsilon \omega_{n}^2}{c^2} \right)  V_n +
\frac{\delta \epsilon \omega_{n}^2}{2 c^2} \left( V_{n+1} +  V_{n-1} \right) \right] =0\nonumber\\
\end{IEEEeqnarray}

The secular equation $V_{n-1}+ b_n V_n +  V_{n+1}=0$ can thus be derived, with
\begin{IEEEeqnarray}{l}
b_n = \frac{2}{\delta}\left(1-\frac{c^2 \beta^2}{\epsilon(\omega_0+n \omega_m)^2}\right)
\label{bn}
\end{IEEEeqnarray}
Now the calculation to obtain the dispersion relation from (\ref{bn}) proceeds similarly to \cite{leakantenna}
in the hypothesis of weak modulation index $\delta<<1$. First the system 
\begin{IEEEeqnarray}{lcr}
V_{-1}+ b_0 V_0 &=& 0 \nonumber\\
b_{-1} V_{-1}+V_{0} &=&0
\label{bnsystem}
\end{IEEEeqnarray}
is written: higher order terms $V_1$ and $V_{-2}$ are neglected because of the properly chosen 
boundary conditions (shortcut filters at $z=0$ and $z=\lambda/4$, at the respective frequencies
$\omega_0+ \omega_m$ and $\omega_0-2 \omega_m$, and also because $V_{-2}$ is higher order in the small modulation index, since it can be shown that $b_{n\neq 0}\sim O(\delta^{-1})$  and $b_{0}\sim O(\delta)$.
Actually, from recursion relations (A-15) and (A-17) in \cite{leakantenna} it appears that the 
n-mode amplitude $V_{n}$ is proportional $\delta^{|n|}$.
From the determinant of (\ref{bnsystem}) we obtain: 
\begin{IEEEeqnarray}{lcr}
b_0 b_{-1} = 1
\label{bndet}
\end{IEEEeqnarray}
Now equation (\ref{bn}) can be plugged into (\ref{bndet}) 
yielding 
\begin{IEEEeqnarray}{rcl}
\left(\beta^2-\frac{\epsilon\omega_0^2}{c^2}\right)\left(\beta^2-\frac{\epsilon(\omega_0-\omega_m)^2}{c^2}\right)=\delta^2\epsilon^2\frac{\omega_0^2 
(\omega_0-\omega_m)^2}{4c^4}
\label{bndetbeta}
\end{IEEEeqnarray}
The biquadratic equation above yields two distinct solutions for $\beta^2$: with no loss of generality we can look for solutions with only positive values of 
$\beta$, since the redundancy due to the presence of a forward and a backward propagating wave is already considered in the $\pm$ sign of (\ref{ansatz}): for 
$\delta=0$ (no modulation) the solutions of (\ref{bndetbeta}) are the usual dispersion relations, $\beta_0=\frac{\sqrt{\epsilon}\omega_0}{c}$ for the fundamental mode and $\beta_{-1}=\frac{\sqrt{\epsilon}(\omega_0-\omega_m)}{c}$ for the low frequency mode.
Solving the biquadratic equation (\ref{bndetbeta}) and keeping the lowest nontrivial order in $\delta$, the dispersion relations follow:
\begin{IEEEeqnarray}{rcl}
\beta_0&=&\frac{\sqrt{\epsilon} \omega_0}{c}+ \frac{\sqrt{\epsilon}\omega_0 (\omega_0-\omega_m)^2 \delta^2 }{8 \omega_m  (2 \omega_0-\omega_m) c} \nonumber \\
\beta_{-1}&=&\frac{\sqrt{\epsilon} (\omega_0-\omega_m)}{c}- \frac{\sqrt{\epsilon}\omega_0^2 (\omega_0-\omega_m) \delta^2 }{8 \omega_m  (2 \omega_0-\omega_m) c}\label{dispersion}   
\end{IEEEeqnarray}
For the two wave numbers $\beta_0$ and $\beta_{-1}$, we obtain from (\ref{bn}) \mbox{$b_0=-\frac{(\omega_0-\omega_m)^2 \delta }{2 \omega_m  (2 \omega_0-\omega_m) }$} 
and \mbox{$b_1=\frac{\omega_0^2 \delta }{2 \omega_m  (2 \omega_0-\omega_m) }$} respectively. Equation (\ref{bnsystem}) yields the low frequency mode amplitude $V_{-1}$ as a function of the fundamental $V_{0}$: the resulting mode combination 
\mbox{$(V_0\exp(-i \omega_0 t)+V_{-1}\exp(-i (\omega_0-\omega_m) t))\exp(\pm i\beta z)$} for the two wave numbers $\beta=\beta_{0,-1}$ solves (\ref{modesV0}). The general voltage solution can be arranged in the following linear combination of four independent elements:
{\small
\begin{IEEEeqnarray}{l}
A \left(\exp(-i \omega_0 t) + \frac{(\omega_0-\omega_m)^2 \delta }{2 \omega_m  (2 \omega_0-\omega_m) }\exp(-i (\omega_0-\omega_m) t)\right) 
\cos(\beta_0 z)+ \nonumber \\
B \left(\exp(-i \omega_0 t) + \frac{(\omega_0-\omega_m)^2 \delta }{2 \omega_m  (2 \omega_0-\omega_m) }\exp(-i (\omega_0-\omega_m) t)\right) 
\sin(\beta_0 z)+  \nonumber \\  
C \left(\exp(-i (\omega_0-\omega_m) t) - \frac{\omega_0^2 \delta }{2 \omega_m  (2 \omega_0-\omega_m) }\exp(-i \omega_0 t)\right) 
\cos(\beta_{-1} z)+ \nonumber \\
D \left(\exp(-i (\omega_0-\omega_m) t) - \frac{\omega_0^2 \delta }{2 \omega_m  (2 \omega_0-\omega_m) }\exp(-i \omega_0 t)\right) 
\sin(\beta_{-1} z)            
\label{Vmodesol}
\end{IEEEeqnarray}
}
The application of the boundary conditions, four independent equations overall, for the two frequencies $\omega_0$ and $\omega_0-\omega_m$ and the two
bounds of the modulated line, completely defines the four coefficients $A,B,C,D$. 
The current can be obtained by \mbox{$I(z,t)=\frac{\partial V(z,t)}{\partial z}\frac{1}{i \omega_n L}$} and is expressed as:
{\small
\begin{IEEEeqnarray}{l}
i\frac{A}{Z_0} \left(\exp(-i \omega_0 t) + \frac{\omega_0 (\omega_0-\omega_m) \delta }{2 \omega_m  (2 \omega_0-\omega_m) }\exp(-i (\omega_0-\omega_m) t)\right) 
\sin(\beta_0 z)+ \nonumber \\
-i\frac{B}{Z_0} \left(\exp(-i \omega_0 t) + \frac{\omega_0 (\omega_0-\omega_m) \delta }{2 \omega_m  (2 \omega_0-\omega_m) }\exp(-i (\omega_0-\omega_m) t)\right) 
\cos(\beta_0 z)+  \nonumber \\  
i\frac{C}{Z_0} \left(\exp(-i (\omega_0-\omega_m) t) - \frac{\omega_0(\omega_0-\omega_m)\delta }{2 \omega_m  (2 \omega_0-\omega_m) }\exp(-i \omega_0 t)\right) 
\sin(\beta_{-1} z)+ \nonumber \\
-i\frac{D}{Z_0} \left(\exp(-i (\omega_0-\omega_m) t) - \frac{\omega_0(\omega_0-\omega_m) \delta }{2 \omega_m  (2 \omega_0-\omega_m) }\exp(-i \omega_0 t)\right) 
\cos(\beta_{-1} z)            
\label{Imodesol}
\end{IEEEeqnarray}
}
where $Z_0=\sqrt{L_{line}/C_{0\ varicap}}=\sqrt{L_{line}/(\epsilon\ C_{line})}$.
The input impedance at $\omega_0$ can be derived by the ratio of the $\omega_0$ components in (\ref{Vmodesol}) and (\ref{Imodesol}) evaluated at $z=0$, that is the feed connection points A1 and A2 in Figure \ref{sideview}:
\begin{IEEEeqnarray}{rcl}
Z_{w0\ in}=Z_0 \frac{A-\frac{\omega_0^2 \delta }{2 (\omega_0^2-\omega_{-1}^2) } C }{-i B + i \frac{\omega_0 \omega_{-1} \delta }{2 (\omega_0^2-\omega_{-1}^2) } D} 
\label{Z0input}
\end{IEEEeqnarray}
where we defined $\omega_{-1}:=\omega_0-\omega_m$.
The antenna input impedance at $\omega_{-1}$, derived by the ratio of the $\omega_0-\omega_m$ components at $z=0$, is: 
\begin{IEEEeqnarray}{rcl}
Z_{w1\ in}=Z_0 \frac{ \omega_{-1}^2 \delta A + 2 (\omega_0^2-\omega_{-1}^2) C}{-i  \omega_0 \omega_{-1} \delta B - i 2 (\omega_0^2-\omega_{-1}^2) D}=- Z_{ant} \nonumber
\label{Z1input}
\end{IEEEeqnarray}
where the minus sign in $- Z_{ant}$ is because the current is conventionally assumed to flow out from the antenna into the transmission line. 
The output impedance at $\omega_0$, the ratio of the $\omega_0$ components in (\ref{Vmodesol}) and (\ref{Imodesol}) evaluated at $z=\pi/(2\beta_0)$, that is at
points B1 and B2 in figure \ref{sideview}, opposite side of the antenna, is:
\begin{IEEEeqnarray}{rcl}
Z_{w0\ out}&=&Z_0 \frac{B-\frac{\omega_0^2 \delta }{2 (\omega_0^2-\omega_{-1}^2) } C
- \frac{\pi}{2}\frac{\omega_0 \omega_{-1} \delta }{2 (\omega_0^2-\omega_{-1}^2) } D}{i A - i \frac{\pi}{2}\frac{\omega_{-1}^2 \delta }{ 2 (\omega_0^2-\omega_{-1}^2) } C +
i \frac{\omega_0 \omega_{-1} \delta }{2 (\omega_0^2-\omega_{-1}^2) } D} 
\label{Z0output}
\end{IEEEeqnarray}
The output impedance at $\omega_{-1}$, the ratio of the $\omega_{-1}$ components evaluated at $z=\pi/(2\beta_0)$, is:
\begin{IEEEeqnarray}{rcl}
Z_{w1\ out}=Z_0 \frac{\frac{\omega_{-1}^2 \delta }{2 (\omega_0^2-\omega_{-1}^2) } B+C
+ \frac{\pi}{2} \frac{ \omega_{-1} }{\omega_0} D}{ i \frac{\omega_0 \omega_{-1} \delta }{2 (\omega_0^2-\omega_{-1}^2) } A + i \frac{\pi}{2}\frac{\omega_{-1}}{ \omega_0} C - i D} 
\label{Z1output}
\end{IEEEeqnarray}
where for both equations (\ref{Z0output}) and (\ref{Z1output}) we approximated $\cos( \frac{\beta_{-1} \pi}{2 \beta_0})\sim1$ and $\sin(\frac{\beta_{-1}\pi}{2 \beta_0})\sim\frac{\omega_{-1} \pi}{2 \omega_0}$. 
Pure imaginary $Z_{w0\ out}$ and $Z_{w1\ out}$ are required in order to ensure that all the energy converted from 
$\omega_0$ to the low mode $\omega_{-1}$ is used at the antenna end, with no energy loss in the filter between B1 and B2 in Figure \ref{sideview}. From inspection of (\ref{Z0output}) and (\ref{Z1output}) this happens if we set $A=\frac{\pi}{2}B\omega_{-1}^2/\omega_0^2$ and $D=0$. From the additional equation (\ref{Z1input}) we obtain:
\begin{IEEEeqnarray}{rcl}
A&=& \frac{2 \pi \omega_{-1} (\omega_{-1}^2 - \omega_0^2 ) Z_0}{(\pi \omega_{-1}^3 Z_0 - 2 i \omega_0^3 Z_{ant} ) \delta}C\nonumber\\
B&=& \frac{4  \omega_0^2 (\omega_{-1}^2 - \omega_0^2 ) Z_0}{\omega_{-1} (\pi \omega_{-1}^3 Z_0 - 2 i \omega_0^3 Z_{ant} ) \delta}C \nonumber\\
D&=&0
\label{ABCD}
\end{IEEEeqnarray}
With these settings the boundary output impedances (\ref{Z0output}) and (\ref{Z1output}) become:
\begin{IEEEeqnarray}{rcl}
Z_{w0\ out}&=&Z_0 \frac{2\omega_0^2}{i\pi\omega_{-1}^2} \nonumber\\
Z_{w1\ out}&=&Z_0 \frac{2\omega_0}{i\pi\omega_{-1}}
\label{Zoutputsimpl}
\end{IEEEeqnarray}
while the input impedance from (\ref{Z0input}) is:
\begin{IEEEeqnarray}{rcl}
Z_{w0\ in}&=& \frac{i \pi \omega_{-1}^2 Z_0}{2 \omega_0^2} + 
             \frac{i \pi \omega_{-1}^4 Z_0 + 2 \omega_0^3 \omega_{-1} Z_{ant}}{8 (\omega_0^2-\omega_{-1}^2)^2} \delta^2\sim 
             Z_{ant} \frac{\omega_{-1}\delta^2}{4 \omega_0} 
\label{Z0inputsimpl}
\end{IEEEeqnarray}

where in the last step we considered  $Z_{ant}>>Z_0$ since it scales as the inverse of the low frequency $\omega_{-1}$.

\section*{Acknowledgment}
This work was fully supported by the patience of dear Stefania, which I want to thank for the English revision of this paper.


\begin{thebibliography}{00}

 
\bibitem{limit1} H. A. Wheeler, "Fundamental limitations of small antennas", 
\emph{Proc. IRE}, vol. 35, pp. 1479-1484, Dec. 1947.

\bibitem{limit2} L. J. Chu, "Physical limitations of omni-directional antennas", 
\emph{J. Appl.Phys.}, vol. 19, pp. 1163-1175, Dec. 1948.

\bibitem{nonfoster1} K.S.Song, "Non-Foster Impedance Matching and Loading Networks for Electrically Small Antennas,"
Ph.D. dissertation, Dept. Elect. and Comp. Engin., The Ohio State University, 2011.



\bibitem{doubleneg1} R. W. Ziolkowski, A.D. Kipple, 
"Application of Double Negative Materials to Increase the Power Radiated by Electrically Small Antennas," \emph{IEEE Trans. on Antennas and Propag.}, vol. 51, no.10, Oct. 2003.

\bibitem{doubleneg2} R. W. Ziolkowski, A.Erentok, "A Metamaterial-Based Efficient Electrically Small Antennas,"
\emph{Trans. IEEE}, vol. AP-54, Jul. 2006, pp. 2113–2130.

\bibitem{doubleneg3} R. W. Ziolkowski, "Metamaterial-Based Antennas: Research and Developments,"
\emph{IEICE Trans. Electron.}, vol. E89–C, no.9, Sep. 2006.

\bibitem{meander}  H.Kanaya, R.K.Pokharel, R.Nabeshima, K.Yoshida, "Design and Performance of an Electrically Small Antenna with Matching Circuit," \emph{Proc. of Asia-Pacific Microwave Conf.}, Bangkok, Dec. 2007.



\bibitem{acoustic1} C.Forrest, "Acoustically driven integrated microstrip antennas and
electromagnetic radiation from piezoelectric devices," Ph.D. dissertation, Dept. Elect. Eng., Iowa State University, 1993.

\bibitem{acoustic2} A.E.Hassanien, M.Breen, M.Li, S.Gong, "A theoretical study of acoustically driven antennas," 
\emph{Journal of Appl. Phys.} 127, 014903 (2020).

\bibitem{acoustic3} A.E.Hassanien, M.Breen, M.Li, S.Gong, "Acoustically Driven and Modulation Inducible Radiating Elements,"
\emph{arXiV: 1906.07797}.

\bibitem{acoustic4} X.Liang, H.Chen, N.Sun, H.Lin, N.Sun, "Novel Acoustically Actuated Magnetoelectric Antennas,"
\emph{Proc. IEEE Internat. Symp. on Antennas and Propag.}, Boston, MA, USA, 2018.

\bibitem{acoustic5} T.Nan et al., "Acoustically actuated ultra-compact NEMS magnetoelectric antennas,"
\emph{Nature Communications} 8(1):296, Dec. 2017.

\bibitem{hadad} L.Goltcman, Y.Hadad, "Scattering from artificial piezoelectriclike meta-atoms and molecules," \emph{Phys. Rev. Lett.} 120(5), 054301, 2018.


\bibitem{switchshort} L. A. Thompson, "Broadband electrically short transmitters via time-varying antenna properties," M.S. thesis, School of Electrical and Computer Engineering, Georgia Institute of Technology, 2017.

\bibitem{switchline} A.Shlivinski, Y.Hadad, "Beyond the Bode-Fano bound: wideband impedance matching for short pulses using temporal switching of transmission line parameters," \emph{Phys. Rev. Lett.}, 121, 204301, 2018.

\bibitem{modulatedC} P.Loghmannia, M.Manteghi, "Broadband Parametric Impedance Matching for Small Antennas Beyond the Bode-Fano Limit," \emph{arXiv: 1907.11683}.



\bibitem{metasurface1} X.Guo, Y.Ding, Y.Duang, X.Ni, "Nonreciprocal metasurface with space–time phase modulation,"
\emph{Light: Science \& Applic.} 8, 123 (2019).

\bibitem{metasurface2} F.Monticone, A.Alù, "Leaky-Wave Theory, Techniques, and Applications: From Microwaves to Visible Frequencies," \emph{Proc. IEEE}, vol. 103, no.5, May 2015.

\bibitem{metasurface3} S.Taravati, "Application of space and time modulated dispersion engineered metamaterials to signal 
processing and magnetless nonreciprocity," Ph.D. dissertation, Dept. G\'enie \'Electr., \'Ecole Polytech. Montr\'eal, 2017.

\bibitem{leakantenna} S.Taravati, C.Caloz, "Mixer-Duplexer-Antenna Leaky-Wave System
Based on Periodic Space-Time Modulation," \emph{IEEE Trans. on Antennas and Propag.}, vol 65, Issue 2, Feb. 2017.

\bibitem{opticaltrans} K.H.Villegas, F.V.Kusmartsev, I.G.savenko, "Optical Transistor for an Amplification of Radiation in a Broadband THz Domain," \emph{Phys. Rev. Lett.} 124, 087701, Feb. 2020.

\bibitem{graphene} X.Wang, A.Diaz Rubio, H.Li, S.Tretyakov, A.Al\`{u}, "Multifunctional Space-Time Metasurfaces," \emph{arXiv: 1910.11812v1}.

\bibitem{MRowe1} J. Manley, H. E. Rowe, "Some general properties of nonlinear elements-part I: General energy relations," \emph{Proc. IRE}, vol. 44, no. 7, pp. 904 – 913, Jul. 1956.

\bibitem{MRowe2}  J. Manley, "Some properties of time varying networks," \emph{IRE Trans. Circ. Theory}, vol. 7, no. 5, pp. 69 – 78, Aug. 1960.

\bibitem{Qfactor}   A.D. Yaghjian, S.R. Best, "Impedance, bandwidth, and Q of antennas," \emph{IEEE Trans. on Antennas and Propag.}, vol 53, Issue 4, Apr. 2005.

\bibitem{Qcomp1}   S.R. Best, "A study of the impedance properties of small antennas," \emph{Proc. Antenna Applications Symposium}, 2007.

\bibitem{Qcomp2}   F.Sarrazin, S.Pflaum, C.Delaveaud "Radiation Efficiency Optimization of Electrically Small Antennas," \emph{Proc. IWAT2016}, Feb. 2016.

\bibitem{shuntC} P.L. Chi, R. Waterhouse, T.Itoh "Antenna Miniaturization Using Slow Wave
Enhancement Factor from Loaded Transmission Line Models," \emph{IEEE Trans. on Antennas and Propag.}, 59(1):48-57, Feb. 2011.


\end{thebibliography}
\end{document}